\documentstyle[12pt]{article}
\def\Re{{\cal R \mskip-4mu \lower.1ex \hbox{\it e}\,}}
\def\Im{{\cal I \mskip-5mu \lower.1ex \hbox{\it m}\,}}

\def\eg{{\it e.g.}}

\def\etal{{\it et al.}}
\def\ibid{{\it ibid}.}
\def\sub#1{_{\lower.25ex\hbox{$\scriptstyle#1$}}}
\def\sul#1{_{\kern-.1em#1}}
\def\sll#1{_{\kern-.2em#1}}
\def\sbl#1{_{\kern-.1em\lower.25ex\hbox{$\scriptstyle#1$}}}
\def\ssb#1{_{\lower.25ex\hbox{$\scriptscriptstyle#1$}}}
\def\sbb#1{_{\lower.4ex\hbox{$\scriptstyle#1$}}}

\def\gev{\,{\rm GeV}}

\def\to{\rightarrow}
\def\mh{\ifmmode m\sbl H \else $m\sbl H$\fi}
\def\mch{\ifmmode m_{H^\pm} \else $m_{H^\pm}$\fi}
\def\mt{\ifmmode m_t\else $m_t$\fi}
\def\mc{\ifmmode m_c\else $m_c$\fi}
\def\mz{\ifmmode M_Z\else $M_Z$\fi}
\def\mw{\ifmmode M_W\else $M_W$\fi}
\def\mws{\ifmmode M_W^2 \else $M_W^2$\fi}
\def\mhs{\ifmmode m_H^2 \else $m_H^2$\fi}
\def\mzs{\ifmmode M_Z^2 \else $M_Z^2$\fi}
\def\mts{\ifmmode m_t^2 \else $m_t^2$\fi}
\def\mcs{\ifmmode m_c^2 \else $m_c^2$\fi}
\def\mchs{\ifmmode m_{H^\pm}^2 \else $m_{H^\pm}^2$\fi}
\def\bsg{\ifmmode b\rightarrow s\gamma \else $b\rightarrow s\gamma$\fi}
\def\ztwo{\ifmmode Z_2\else $Z_2$\fi}
\def\zone{\ifmmode Z_1\else $Z_1$\fi}
\def\mtwo{\ifmmode M_2\else $M_2$\fi}
\def\mone{\ifmmode M_1\else $M_1$\fi}
\def\tb{\ifmmode \tan\beta \else $\tan\beta$\fi}
\def\xw{\ifmmode x\sub w\else $x\sub w$\fi}
\def\ch{\ifmmode H^\pm \else $H^\pm$\fi}
\def\lum{\ifmmode {\cal L}\else ${\cal L}$\fi}
\def\inpb{\ifmmode {\rm pb}^{-1}\else ${\rm pb}^{-1}$\fi}
\def\infb{\ifmmode {\rm fb}^{-1}\else ${\rm fb}^{-1}$\fi}
\def\epem{\ifmmode e^+e^-\else $e^+e^-$\fi}
\def\ppb{\ifmmode \bar pp\else $\bar pp$\fi}

\newskip\zatskip \zatskip=0pt plus0pt minus0pt
\def\matth{\mathsurround=0pt}

\def\gsim{\mathrel{\mathpalette\atversim>}}
\def\atversim#1#2{\lower0.7ex\vbox{\baselineskip\zatskip\lineskip\zatskip
  \lineskiplimit 0pt\ialign{$\matth#1\hfil##\hfil$\crcr#2\crcr\sim\crcr}}}

\renewcommand{\thefootnote}{\fnsymbol{footnote}}

\begin{document} \begin{titlepage}
\setcounter{page}{1}
\thispagestyle{empty}
\rightline{\vbox{\halign{&#\hfil\cr
&ANL-HEP-PR-93-37\cr
&May 1993\cr}}}
\vspace{1in}
\begin{center}

{\Large\bf
Using $b \to s\gamma$ to Probe Top Quark Couplings}
\footnote{Research supported by the
U.S. Department of
Energy, Division of High Energy Physics, Contract W-31-109-ENG-38.}
\medskip

\normalsize JOANNE L. HEWETT and  THOMAS G. RIZZO
\\ \smallskip
High Energy Physics Division\\Argonne National
Laboratory\\Argonne, IL 60439\\

\end{center}

\begin{abstract}

Possible anomalous couplings of the top-quark to on-shell photons and gluons
are constrained by the recent results of the CLEO Collaboration on both
inclusive and exclusive radiative $B$ decays.  We find that the process \bsg\
can lead to reasonable bounds on both the anomalous electric and
magnetic dipole moments of the top-quark, while essentially no limits are
obtained on the corresponding chromoelectric and chromomagnetic moments,
which enter the expression for the decay rate
only through operator mixing.

\end{abstract}

\renewcommand{\thefootnote}{\arabic{footnote}} \end{titlepage}


The Standard Model (SM) of electroweak interactions is in very good agreement
with present experimental data\cite{rolandi}.  Nonetheless, it is believed
to leave many questions unanswered, and this belief has resulted in numerous
theoretical and experimental attempts to discover a more fundamental
underlying theory.  Various types of experiments may expose the existence of
physics beyond the SM, including the search for direct production of exotic
particles at high-energy colliders.  A complementary approach in hunting for
new physics is to examine its indirect effects in higher order processes.
For example, even though the top-quark has yet to be discovered, it has already
made its presence known
through loop order processes, such as rare decays of the b-quark.
Since the top-quark is far more massive than the other SM fermions, its
interactions may be quite sensitive to new physics originating at a higher
scale.
If there are any deviations from SM expectations in the properties of the
top-quark, they may indirectly lead to modifications in the anticipated
branching fractions for these one-loop induced b-quark decays.

The possibility of having top-quarks with anomalous couplings to various gauge
bosons has been discussed in the literature\cite{soni,nlc}.  Strong bounds can
be placed on these anomalous couplings at future colliders, such as the
SSC/LHC\cite{soni} and the next linear $e^+e^-$ linear collider
(NLC)\cite{nlc},
which rely on direct production of top-quark pairs.  Since the t-quark has yet
to appear at the Tevatron, it is clear that any
restrictions on these couplings at present can only be obtained indirectly.
In this paper we examine the effects of anomalous couplings of the
top-quark to on-shell photons and gluons on the process
\bsg.  If the t-quark has large anomalous couplings, then the resulting
prediction of the rate for \bsg\ would conflict with experiment.
The CLEO collaboration has recently\cite{thorn} observed the exclusive
decay $B\to K^*\gamma$ with a branching fraction of $(4.5\pm 1.5\pm 0.9)
\times 10^{-5}$ and has placed an upper limit on the inclusive quark-level
process \bsg\ of $B(\bsg)< 5.4\times 10^{-4}$ at 95$\%$ CL. Using a
conservative estimate
of the ratio of exclusive to inclusive decay rates\cite{sonilatt}, the
observation of the exclusive process implies the lower bound
$B(\bsg)>0.6\times 10^{-4}$ at 95$\%$ CL.  These values for the branching
fractions are consistent with expectations from the SM\cite{smbsg}.

The most general form of the Lagrangian which describes the interactions of
top-quarks to on-shell photons, assuming operators of dimension-five or less
only, is
\begin{equation}
{\cal L} = e\bar t \left[ Q_t\gamma_\mu+{1\over 2m_t}\sigma_{\mu\nu}
(\kappa_\gamma +i\tilde\kappa_\gamma \gamma_5)q^\nu \right] t A^\mu \,,
\end{equation}
where, for simplicity, we have assumed that the ordinary dimension-four
interaction is parity conserving.  Here, $Q_t$ is the electric charge of
the t-quark, $\kappa_\gamma\ (\tilde\kappa_\gamma)$ is the anomalous magnetic
(electric) dipole moment, and \mt\ represents the mass of the top-quark.
A similar expression is obtained for the interactions of the t-quark with
gluons with obvious substitutions in the above.  To simplify our analysis, we
will also assume that only one of either the electric or magnetic dipole
operators is non-zero. Clearly, if all four operators are non-vanishing, their
separate contributions would be quite impossible to disentangle using an
analysis of the \bsg\ rate alone.

Our investigation of this process proceeds as follows.  To obtain the \bsg\
branching fraction, the inclusive $b \to s\gamma$ rate is scaled to that of
the semileptonic decay $b\to X\ell\nu$.  This removes major
uncertainties in the calculation associated with ($i$) an overall factor of
$m_b^5$ which appears in both expressions and ($ii$) the imprecisely known
Cabbibo-Kobayashi-Maskawa (CKM) factors.   We use the latest data
on the semileptonic branching fraction{\cite {pdg,drell}}, which is given by
$B(b \to X \ell \nu)=0.108$, to rescale our result.
The semileptonic rate is calculated including both charm and non-charm modes,
assuming $|V_{ub}/V_{cb}|=0.1$, and includes both phase space and QCD
correction
s
with $m_b=5\gev$ and $m_c=1.5\gev${\cite {cab}}.   The calculation of
$\Gamma(\bsg)$ employs the next-to-leading log evolution equations
for the coefficients of the operators in the effective Hamiltonian due to
Misiak{\cite {misiak}}, the gluon bremsstrahlung corrections of Ali and
Greub{\cite {ali}}, the $m_{top} \neq M_W$ corrections of Cho and Grinstein
{\cite {cho}}, a running $\alpha_{QED}$ evaluated at the b-quark mass scale,
and 3-loop evolution of the running $\alpha_s$ matched to the value obtained
at the $Z$ scale via a global analysis{\cite {ellis}} of all data. Phase space
corrections for the strange quark mass in the final state are included and the
ratio of CKM mixing matrix elements in the scaled decay rate,
$|V_{tb}V_{ts}/V_{cb}|$, is taken to be unity.
The details of this procedure are
presented elsewhere{\cite {jlh}}. To complete
the calculation we use the one-loop matching conditions for the
various operators{\cite {misiak}} in a form that includes contributions from
both the SM and the top-quark anomalous couplings.

In practice, only the coefficients
of the dipole $b\to s$ transition operators, traditionally denoted as
$O_7$ and $O_8$, are modified by the presence of the anomalous couplings.
At the $W$ scale, $O_7$ is the only operator which mediates the decay \bsg,
however, mixing occurs between the various $b\to s$ transition operators
during the evolution of the coefficient of $O_7$ to the b-quark mass scale,
so that in principle all the operators can contribute at the scale $m_b$.
We can write the coefficients of $O_7$ and $O_8$ at the $W$ scale as,
\eg\,
\begin{eqnarray}
c_7(M_W) &=& G^7_{SM}+\kappa_\gamma G_1+i\tilde\kappa_\gamma G_2 \,, \\
c_8(M_W) &=& G^8_{SM}+\kappa_g G_1+i\tilde\kappa_g G_2 \,, \nonumber
\end{eqnarray}
with
\begin{eqnarray}
G^7_{SM} &=& -{1\over 2} \left[ {-3x^3+2x^2\over 2(1-x)^4}\ln x
-{8x^3+5x^2-7x\over 12(1-x)^3} \right] \,, \nonumber \\
G^8_{SM} &=& -{1\over 2} \left[ {3x^2\ln x\over 2(x-1)^4} +
{x^3-5x^2-2x\over 4(x-1)^3} \right] \,, \nonumber \\
G_1 &=& {1\over x-1} - {\ln x\over (x-1)^2} + { {1\over 2}x-1\over (x-1)^3}
\left[ {1\over 2}x^2-2x+{3\over 2}+\ln x\right] - {1\over 4} \,, \\
G_2 &=& {1\over x-1} - {\ln x\over (x-1)^2} - {1\over 4} \,, \nonumber
\end{eqnarray}
where $x=m_t^2/M_W^2$.  Note that this result is completely finite
and that we do not have to resort to any use of cut-offs to analyze our
results.

In Fig.~1a we show the predicted \bsg\ branching fraction for several different
top-quark masses assuming {\it {only}} the anomalous magnetic dipole moment of
the top is non-zero. For large negative(positive) values of $\kappa_\gamma$,
we see that the branching fraction exceeds the inclusive CLEO upper(lower)
bound.  While the constraint on $\kappa_\gamma$ from the CLEO upper limit does
not appear to be sensitive to $m_t$, the restriction on $\kappa_\gamma$ from
the lower CLEO bound
varies significantly for $m_t$ in the range $120 - 200$ \gev . While
the anomalous magnetic dipole moment effects the $O_7$ operator directly, the
anomalous chromomagnetic dipole moment only contributes to the rate for \bsg\
indirectly through operator mixing. Thus we would naively expect that the
resultant bounds that are obtainable on this parameter to be quite weak. We see
this quite explicitly in Fig.~1b where we assume that only $\kappa_g$ is
non-zero. It is clear that unless extraordinarily large values of $\kappa_g$
are realized, the present \bsg\ data does not constrain this parameter. If
{\it {both}} $\kappa_g$ and $\kappa_\gamma$ are taken to be non-zero,
the bounds obtainable
from Fig.~1a on $\kappa_\gamma$ will not be significantly modified unless huge
values of $\kappa_g$ are assumed. To demonstrate this we show in Fig.~1c the
allowed range of $\kappa_\gamma$ at the 95$\%$ CL as a function of $m_t$,
assuming that $\kappa_g$ is absent (solid curve) or is identical in value to
$\kappa_\gamma$ (dashed curve). We see that both sets of constraints
are remarkably similar. The general weakening of the limits with increasing
$m_t$ should be noted.

In the case of a non-zero anomalous electric or chromoelectric dipole moment,
a somewhat different situation occurs due to the relative phase between these
and the conventional SM contributions to $c_{7,8}(M_W)$. When evolved down
to the b-quark scale, these contributions do not interfere with those of the
SM and thus can only appear quadratically in the modified expression for the
\bsg\ rate. Thus, we will assume both $\tilde\kappa_\gamma$ and
$\tilde\kappa_g$ to be positive semi-definite in our numerical analysis.
Since the contribution from these anomalous
couplings can only {\it {increase}} the prediction
to the \bsg\ rate over that given by the SM, we
anticipate that only the CLEO upper bound will provide a constraint. This
expectation is borne out by our explicit calculations.

Fig.~2a displays the \bsg\ branching fraction as a function of
$\tilde\kappa_\gamma$ for several different top-quark masses assuming all
other anomalous couplings are zero. Given the set of
top-quark masses we consider, the bound on this parameter apparently
strengthens as the value of $m_t$ increases. This conclusion is quite valid
provided $m_t\gsim 130$ \gev, however, as we will see below,
it is not quite correct for smaller values of $m_t$.
As in the case of the chromomagnetic dipole moment, any constraints
on the chromoelectric dipole moment are expected to be quite poor as it
only enters
into the expression for the \bsg\ decay rate through operator mixing. Fig.~2b
shows this is indeed the case, as there is apparently very little sensitivity
to $\tilde\kappa_g$ alone even for very large values of this anomalous
coupling.  In Fig.~2c we present the $m_t$ dependence of the 95$\%$ CL upper
bound on $\tilde\kappa_\gamma$ both when all other anomalous couplings are
absent (solid) as well as when $\tilde\kappa_g=\tilde\kappa_\gamma$ (dashed).
Again, very little difference is
seen between the two cases, demonstrating the lack of sensitivity of
\bsg\ to the chromoelectric moment of the top. As pointed out above, the bound
on $\tilde\kappa_\gamma$ strengthens with increasing $m_t$ for values in excess
of 130 \gev . However, we see that for smaller values of $m_t$ below 130 GeV
the limits also become stronger.  This is due to a cancellation between the
various terms in $G_2$ near $m_t=130$ \gev .

How will the constraints we have obtained be improved in the future? From \bsg\
itself, we see that any imaginable improvement in the data will not
qualitatively alter the allowed ranges we have obtained unless the top-quark
mass is known. Even in this case, the other calculational uncertainties
render it unlikely that drastic improvements are possible from this process
alone. Of course, input from other processes involving top-quark loops may
be of some help and should be aggressively investigated.
Clearly, the next major step forward will be the examination of the top-quark
production process itself after the t-quark is found. Both the SSC/LHC and the
NLC will be able to probe anomalous couplings which are two to three orders
of magnitude smaller than those discussed here. We remind the reader, however,
that the anomalous couplings that can be examined at these colliders will be
for {\it {on-shell}} top-quarks with {\it {off-shell}} photons and
gluons, \eg , the situation opposite to that which we examine here.

In summary, we have shown that the new CLEO results on
radiative $B$ decays place strong constraints on the anomalous electric and
magnetic dipole moment couplings of the top-quark even though it has not yet
been directly observed at the Tevatron. The corresponding limits we obtain on
the chromoelectric and chromomagnetic dipole moments are quite weak as they
enter our calculation only via operator mixing.  Clearly other low energy
processes might also lead to constraints on such anomalous couplings and
should be examined. It is most likely, however, that we will have to wait for
detailed studies of top-quark production at future colliders before more
restrictive bounds can be obtained.

\vskip.25in
\centerline{ACKNOWLEDGEMENTS}

The authors would like to thank A.\ Soni, E.\ Thorndike, Y.\ Rosen, and
S.\ Pakvasa  for discussions related to this work. The authors would also
like to thank the High Energy Physics group at the University of Hawaii, where
this work was completed, for its hospitality and use of its facilities. This
research was supported in part by the U.S.~Department of Energy
under contract W-31-109-ENG-38.

\newpage

%
\def\MPL #1 #2 #3 {Mod.~Phys.~Lett.~{\bf#1},\ #2 (#3)}
\def\NPB #1 #2 #3 {Nucl.~Phys.~{\bf#1},\ #2 (#3)}
\def\PLB #1 #2 #3 {Phys.~Lett.~{\bf#1},\ #2 (#3)}
\def\PR #1 #2 #3 {Phys.~Rep.~{\bf#1},\ #2 (#3)}
\def\PRD #1 #2 #3 {Phys.~Rev.~{\bf#1},\ #2 (#3)}
\def\PRL #1 #2 #3 {Phys.~Rev.~Lett.~{\bf#1},\ #2 (#3)}
\def\RMP #1 #2 #3 {Rev.~Mod.~Phys.~{\bf#1},\ #2 (#3)}
\def\ZP #1 #2 #3 {Z.~Phys.~{\bf#1},\ #2 (#3)}
\def\IJMP #1 #2 #3 {Int.~J.~Mod.~Phys.~{\bf#1},\ #2 (#3)}

\newpage

%
{\bf Figure Captions}
\begin{itemize}

\item[Figure 1.]{The branching fraction for $b \to s\gamma$ as a function of
(a) $\kappa_\gamma$ with $\kappa_g=0$ or (b) $\kappa_g$ with $\kappa_\gamma=0$,
assuming $m_t$=120(140, 160, 180, 200)\gev ~corresponding to the dotted(dashed,
dash-dotted, solid, square-dotted) curve. The solid horizontal lines are the
95$\%$ CL upper and lower bounds from CLEO. (c) The allowed range of
$\kappa_\gamma$ as a function of $m_t$ assuming $\kappa_g=0$ (solid curve) or
$\kappa_g=\kappa_\gamma$ (dashed curve).}
\item[Figure 2.]{Same as Fig.~1 but for the couplings $\tilde\kappa_g$ and
$\tilde\kappa_\gamma$.}
\end{itemize}

\end{document}